\documentclass[12pt]{article}
\usepackage{amsfonts}
\usepackage{mathrsfs}
\usepackage{tabu}
\usepackage{amsfonts,amssymb,eucal,amsmath,epsfig}
\title{ On the symmetric central configurations for the planar 1+4-body problem}
\author{{\small\sc Chunhua Deng$^1$ and Shiqing  Zhang$^2$}\\
\small 1. Faculty of Mathematics and Physics, Huaiyin
Institute of
Technology, Huai'an 223003, China\\
\small 2. College of Mathematics, Sichuan University, Chengdu 610064, China\\
}
\date{}

\begin{document}
\maketitle

{\bf Abstract:}
We study the planar symmetric central configurations of the $1+4$-body problem where the symmetry axis does not contain any infinitesimal masses.
Under certain assumptions we find analytically some central configurations, and also get some numerical results of symmetric central configurations where infinitesimal masses are not necessarily equal.


AMS Subject Classification 34C15, 34C25

\section*{1. Introduction}
\setcounter{section}{1} \setcounter{equation}{0}

A very old problem in Celestial Mechanics is the study of central configurations for the $n$-body problem.
One of the reasons why central configurations are interesting is that they allow us to
construct exact solutions of the $n$-body problem. Central configurations also have other interesting
properties in the study of the $n$-body problem, see [1,2,8-12,15,17,21-24] for
details.

In this paper we consider a restricted version of the problem of planar central
configurations, i.e., we study the limit case of one large mass and $N$ small
masses as the small masses tend to zero (planar $1+N$-body problem).
This problem was first considered by Maxwell [13] trying to construct a model
for Saturn's rings.
Hall [7] studied the planar central configuration of
the $1+N$-body problem where the $N$ small masses are equal.
He found that, when N is sufficiently large, the only possible
relative equilibrium is Maxwell's ring, that is, a regular $N$-gon with a central mass,
and that other configurations are possible for small $N$. Moeckel [14] found a necessary
and sufficient condition for the linear stability of relative equilibria of the $1+N$-
body problem with $N$ small but not necessarily equal masses.
Recently these configurations have attracted the attention
of astronomers. Renner and Sicardy [18] suggest that the presence of coorbital satellites might explain, at
least partly, the confinement of Neptune¡¯s ring arcs.
Corset al. [6] proved that there are only three symmetric central configuration of
the $1+4$-body problem with four separate identical satellites. Albouy and Fu [2] proved that all central configurations
of the $1+4$-body problem are symmetric which settles the question in this case.
A. Oliveira and H. Cabral [16] showed that, for the planar $1+4$-body problem where the satellites have different infinitesimal masses and two of them are diametrically opposite in a circle, the configurations are necessarily symmetric and the other satellites have the same mass. Moreover they prove that the number of central configurations in this case is in general one, two or three and, in the special case where the satellites diametrically opposite have the same mass, they prove that the number of central configurations is one or two and give the exact value of the ratio of the masses that provides this bifurcation. Many other results can be found in [3,5,17,19,20].
Here we study the planar symmetric central configurations of the $1+4$-body problem with $\theta_1=\theta_3$, i.e. the symmetry axis does not contain any infinitesimal masses, where the satellites may have different infinitesimal masses.

\section*{2. Preliminaries}
\setcounter{section}{2} \setcounter{equation}{0}
Consider $n$ particles of masses $m_1, \cdots, m_n$ in $\mathbb{R}^2$ subject to their mutual Newtonian gravitational interaction. In an inertial reference frame and choosing appropriate units, the equations of motion are

\begin{equation}
m_i\ddot{q}_i=\frac{\partial U(q)}{\partial q_i}, i=1,2,\cdots,n
\end{equation}
where

$$U(q)=U(q_1,q_2,\cdots,q_n)=\sum\limits_{1\leq k<j\leq
n}\frac{m_km_j}{|q_k-q_j|}$$
is the Newtonian potential of system (2.1). The position vector $q
=(q_1,q_2,\cdots,q_n)\in (\mathbb{R}^2)^n$ is often referred to the
configuration of the system.

 A configuration $q =(q_1,\cdots,q_n)\in
(\mathbb{R}^2)^n$ with $\sum_{i=1}^{n} m_iq_i=0$ is called a \textit{central
configuration} if there exists some positive constant $\lambda$,
called Lagrangian multiplier, such that
\begin{equation}
-\lambda q_i=\sum\limits_{j=1,j\neq
i}^{n}\frac{m_j(q_j-q_i)}{|q_j-q_i|^3},\quad i=1,2,\cdots,n.
\end{equation}

We are interested in the planar $n=1+N$ body problem, where the big mass is
equal to 1 with position $q_0=0$. The remaining $N$ bodies with positions $q_i$, called
satellites, have masses $m_i=\mu_i\epsilon, i=1,2,\cdots,N$, where $\mu _i\in \mathbb{R}^+$
and $\epsilon>0$ is a small parameter that tends to zero.
In all central configuration of the planar $1+N$-body problem the satellites lie on
a circle centered at the big mass ([4]), i.e. the satellites are co-orbital.

We exclude collisions in the definition of central configuration and take the angles $\theta_i$
between two consecutive particles as coordinates, we refer to [2,6] for details. In
these coordinates the space of configuration is the simplex

$$\bigtriangleup=\{\theta=(\theta_1,\theta_2,\cdots,\theta_N): \sum_{i=1}^{N}\theta_i=2\pi, \theta_i>0, i=1,2,\cdots,N\}$$
and the equations characterizing the central configurations of the planar $1+N$-body
problem are
\begin{equation}
\begin{aligned}
 \mu_2&f(\theta_1)+\mu_3f(\theta_1+\theta_2)+\cdots+\mu_Nf(\theta_1+\cdots+\theta_{N-1})=0,\\
\mu_3&f(\theta_2)+\mu_4f(\theta_2+\theta_3)+\cdots+\mu_1f(\theta_2+\cdots+\theta_N)=0,\\
\mu_4&f(\theta_3)+\mu_5f(\theta_3+\theta_4)+\cdots+\mu_2f(\theta_3+\cdots+\theta_N+\theta_1)=0,\\
\cdots&\\
\mu_N&f(\theta_{N-1})+\mu_1f(\theta_{N-1}+\theta_N)+\cdots+\mu_{N-2}f(\theta_{N-1}+\theta_N+\theta_1+\cdots+\theta_{N-3})=0,\\
\mu_1&f(\theta_N)+\mu_2f(\theta_N+\theta_1)+\cdots+\mu_{N-1}f(\theta_N+\theta_1+\cdots+\theta_{N-2})=0,\\
\theta_1&+\cdots+\theta_N=2\pi,\\
 \end{aligned}
 \end{equation}
 where $f(\theta)=sin(\theta)\left(1-\frac{1}{8|sin^3(\frac{\theta}{2})|}\right).$

 In the case of four satellites system (2.3) is
 \begin{equation}
\begin{aligned}
 \mu_2&f(\theta_1)+\mu_3f(\theta_1+\theta_2)+\mu_4f(\theta_1+\theta_2+\theta_3)=0,\\
\mu_3&f(\theta_2)+\mu_4f(\theta_2+\theta_3)+\mu_1f(\theta_2+\theta_3+\theta_4)=0,\\
\mu_4&f(\theta_3)+\mu_1f(\theta_3+\theta_4)+\mu_2f(\theta_3+\theta_4+\theta_1)=0,\\
\mu_1&f(\theta_4)+\mu_2f(\theta_4+\theta_1)+\mu_3f(\theta_4+\theta_1+\theta_2)=0,\\\
\theta_1&+\theta_2+\theta_3+\theta_4=2\pi.\\
 \end{aligned}
 \end{equation}

The function $f(\theta)$ defined above plays a key role in this problem (Figure 1).
The following two lemmas state some properties of $f$ and its derivatives which will be
used to prove our results.
Lemma 2.1 can be found in [16] and Lemma 2.2 can be proved straightforwardly.

\begin{figure}[htb]
\begin{center}
\includegraphics[width=11cm,height=7cm]{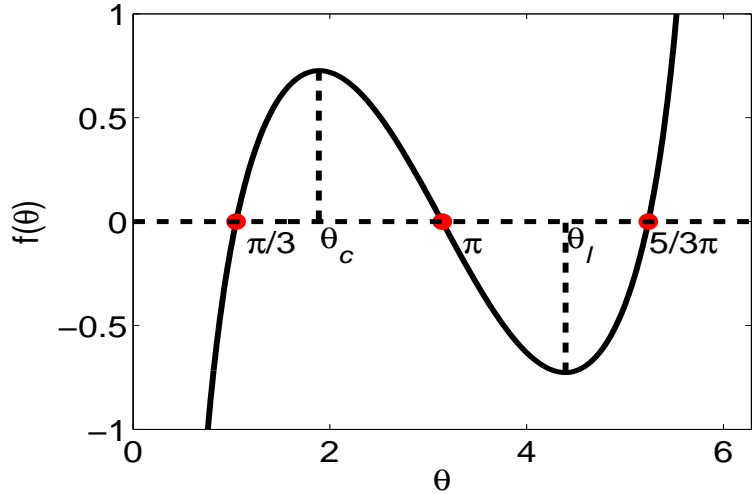}
\vspace{0cm}\caption{\label{TheLabel}\footnotesize Graph of f}
\end{center}
\end{figure}

\noindent{\bf Lemma 2.1.} The function $f(\theta)$
satisfies:\\
1)$f(\pi/3)=f(\pi)=f(5\pi/3)=0;$\\
2) $f(\pi-\theta)=-f(\pi+\theta), f(2\pi-\theta)=-f(\theta),  \forall \theta\in(0,\pi);$\\
3)$f'(\theta)=cos(\theta)+\frac{3+cos(\theta)}{16sin^3(\theta/2)}\geq f'(\pi)=-7/8, \text{for all  }  \theta\in (0,2\pi);$\\
4)In $(0,\pi)$ there is a unique critical point $\theta_c$ of $f$,
where $3\pi/5<\theta_c<2\pi/3$, such that $f'(\theta)>0$ in $(0,\theta_c)$, $f'(\theta)<0$ in $(\theta_c,\pi)$;
In $(\pi,2\pi)$ there is a unique critical point $\theta_l=2\pi-\theta_c$ of $f$,
where $4\pi/3<\theta_l<7\pi/5$, such that $f'(\theta)>0$ in $(\theta_l,2\pi)$, $f'(\theta)<0$ in $(\pi,\theta_l)$.

\noindent{\bf Lemma 2.2.}
$f''(\theta)=-sin(\theta)-\frac{(11+cos\theta)cos(\theta/2)}{32 sin^4(\theta/2)}$, and \\
1)$f''(\theta)<0$ in $(0,\pi)$, $f''(\theta)>0$ in $(\pi,2\pi)$;\\
2)$f'(\theta)<0$ in $(\theta_c,\theta_l)$, $f'(\theta)>0$ in $(0,\theta_c)\cup (\theta_l, 2\pi)$;\\
3)$f(\theta)<0$ in $(0,\pi/3)\cup(\pi,5\pi/3)$, $f(\theta)>0$ in $(\pi/3,\pi)\cup(5\pi/3,2\pi)$.\\

A coorbital central configuration $(\theta_1,\theta_2,\cdots,\theta_N)$ of the planar
 $1+N$-body problem is symmetric with respect to a straight line $L$ containing
 the central body, if modulus a cyclic permutation of the angles we have,

 1. when $N$ is even either
 $$\theta_1=\theta_N, \theta_2=\theta_{N-1}, \cdots,\theta_{\frac{N}{2}}=\theta_{\frac{N+2}{2}},$$
in this case the symmetry axis $L$ contains two satellites, or
$$\theta_1=\theta_{N-1}, \theta_2=\theta_{N-2}, \cdots,\theta_{\frac{N-2}{2}}=\theta_{\frac{N+2}{2}},$$
in this case the symmetry axis $L$ contains no satellites;

\begin{figure}[htb]
\begin{center}
\includegraphics[width=12cm,height=9cm]{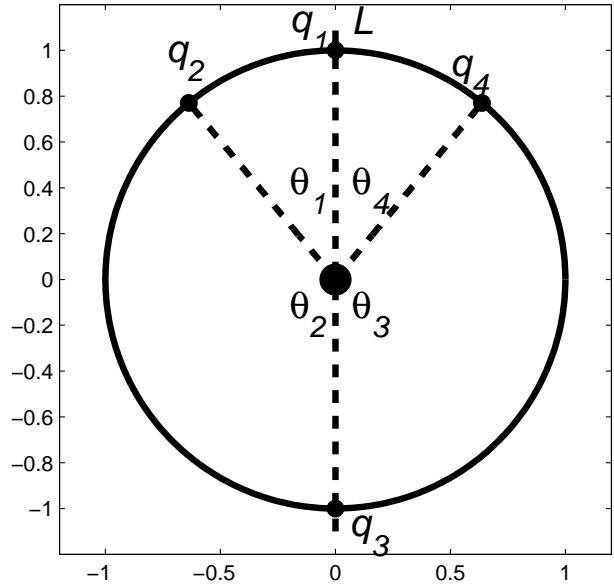}
\vspace{0cm}\caption{\label{TheLabel}\footnotesize The symmetric configuration with $\theta_1=\theta_4$}
\end{center}
\end{figure}

\begin{figure}[htb]
\begin{center}
\includegraphics[width=12cm,height=9cm]{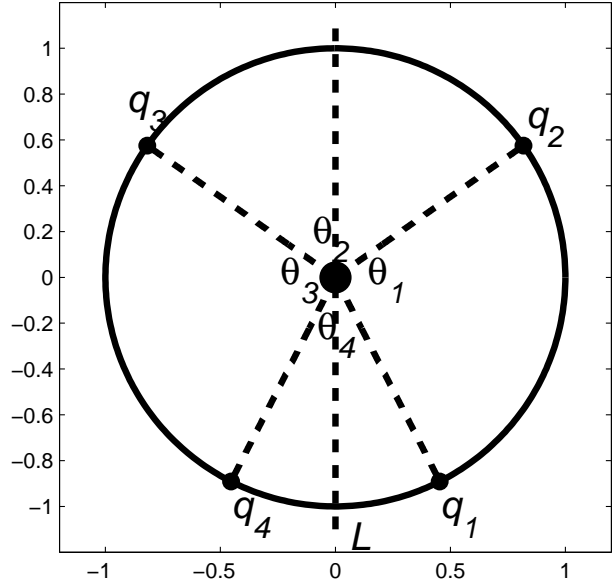}
\vspace{0cm}\caption{\label{TheLabel}\footnotesize The symmetric configuration with $\theta_1=\theta_3$}
\end{center}
\end{figure}

2. and when $N$ is odd,
 $$\theta_1=\theta_N, \theta_2=\theta_{N-1}, \cdots,\theta_{\frac{N-1}{2}}=\theta_{\frac{N+3}{2}},$$
 in this case the symmetry axis $L$ contains one satellite.
In the case of $1+4$-body problem, the symmetric central configuration contains $\theta_1=\theta_3$ or $\theta_1=\theta_4$. Oliveira and Cabral [16] have completed the first case $\theta_1=\theta_4$, where the symmetry axis contains $q_1$ and $q_3$, and $\theta_2=\theta_3$. We consider the symmetric central configuration of $1+4$-body problem with $\theta_1=\theta_3$, where the symmetry axis does not contain any satellites, and $\theta_2$ $\theta_4$ are not necessarily equal. This case are more complicated than the former one.

\section*{3. Main results}
\setcounter{section}{3} \setcounter{equation}{0}

 For the the symmetric central configuration of $1+4$-body problem, we consider now the case $\theta_1=\theta_3$.
 Using the property of $f$ that $f(2\pi-x)=-f(x)$ then system (2.4) becomes
 \begin{equation}
\begin{aligned}
 \mu_2&f(\theta_1)+\mu_3f(\theta_1+\theta_2)-\mu_4f(\theta_4)=0,\\
\mu_3&f(\theta_2)+\mu_4f(\theta_1+\theta_2)-\mu_1f(\theta_1)=0,\\
\mu_4&f(\theta_1)-\mu_1f(\theta_1+\theta_2)-\mu_2f(\theta_2)=0,\\
\mu_1&f(\theta_4)-\mu_2f(\theta_1+\theta_2)-\mu_3f(\theta_1)=0,\\
\theta_1&+\theta_2+\theta_3+\theta_4=2\pi,\\
 \end{aligned}
 \end{equation}
 where $0<\theta_2,\theta_4<2\pi, 0<\theta_1=\theta_3<\pi$.

 Let
$$
 A=\left(\begin{array}{cccc}
    \mu_2 & 0 & -\mu_4 & \mu_3\\
     -\mu_1 & \mu_3 & 0 & \mu_4\\
    \mu_4 & -\mu_2& 0 & -\mu_1\\
     -\mu_3 & 0 &\mu_1 & -\mu_2\\
 \end{array}\right),
 X=\left(       \begin{array}{ccc}
  f(\theta_1)\\
   f(\theta_2)\\
   f(\theta_4)\\
  f(\theta_1+\theta_2) \\
      \end{array}\right),$$
then the system (3.1) can be rewritten as
$$AX=0.$$
With simple computation, we have $det(A)=0$ for any $\mu_1,\mu_2,\mu_3,\mu_4$.

\noindent{\bf Lemma 3.1.} Let $(\theta_1,\theta_2,\theta_3,\theta_4)$ be a symmetric coorbital central configuration solution of system (2.4) with $\theta_1=\theta_3$, then for $f(\theta_i), i=1,2,4$ and $f(\theta_1+\theta_2)$, there can  be at most one to be zero.

 {\bf Proof.}  First we prove that $f(\theta_i), i=1,2,4$ and $f(\theta_1+\theta_2)$ can not to be zero simultaneously.
 Since $f(\theta_1)=0$, and $0<\theta_1=\theta_3<\pi$, then $\theta_1=\theta_3=\pi/3$.
By $f(\theta_1+\theta_2)=0$, we have $\theta_1+\theta_2=\pi$ or $5\pi/3$.

If $\theta_1+\theta_2=\pi$, then $\theta_2=2\pi/3$ and $\theta_4=2\pi/3$, which is a contradiction with the fact $f(\theta_4)=0$.

If $\theta_1+\theta_2=5\pi/3$, then $\theta_2=4\pi/3$ and $\theta_4=0$ which have contradiction with $\theta_4>0$.

Moreover when some two of $f(\theta_1+\theta_2)$ and $f(\theta_i), i=1,2,4$ equal to zero, with simple calculation, we obtain that they are all equal to zero. Thus we complete the proof.

 By lemma 3.1, we consider the following cases that one of $f(\theta_1+\theta_2)$ and $f(\theta_i), i=1,2,4$ equals to zero while the others are not zero, and also the case that all of $f(\theta_i)$, $i=1,2,4$ and $f(\theta_1+\theta_2)$ are not zero.

$f(\theta_1+\theta_2)=0$ implies $\theta_1+\theta_2=\pi/3,\pi$ or $5\pi/3$. We get the following three Theorems:

\noindent{\bf Theorem 3.2.}  Let $(\theta_1,\theta_2,\theta_3,\theta_4)$ be a coorbital central configuration solution of system (2.4).
Suppose that $\theta_1=\theta_3$, $\theta_1+\theta_2=\pi/3$. If $\mu_1\mu_2\neq\mu_3\mu_4$, there exists no central configuration. If $\mu_1\mu_2=\mu_3\mu_4$, there is exactly one central configuration $(\theta_0, \pi/3-\theta_0, \theta_0, 5\pi/3-\theta_0)$ in the $1+4$ body problem, where $\theta_0\approx 0.6281$ is the unique root of the following equation in interval $(0,\pi/3)$
$$f^2(\theta)-f(\pi/3-\theta)f(5\pi/3-\theta)=0.$$
Moreover, $\frac{\mu_1}{\mu_3}=\frac{\mu_4}{\mu_2}=\frac{f(\pi/3-\theta_0)}{f(\theta_0)}$.

\noindent{\bf Theorem 3.3.}  Let $(\theta_1,\theta_2,\theta_3,\theta_4)$ be a coorbital central configuration solution of system (2.4). Suppose that $\theta_1=\theta_3$, $\theta_1+\theta_2=\pi$. If $\mu_1\mu_2\neq\mu_3\mu_4$, there exists no central configuration. If $\mu_1\mu_2=\mu_3\mu_4$, there is exactly one central configuration $(\pi/2, \pi/2, \pi/2, \pi/2)$ in the $1+4$ body problem. Moreover,
$\mu_1=\mu_3, \mu_2=\mu_4$.

\noindent{\bf Theorem 3.4.}  Let $(\theta_1,\theta_2,\theta_3,\theta_4)$ be a coorbital central configuration solution of system (2.4).
Suppose that $\theta_1=\theta_3$, $\theta_1+\theta_2=5\pi/3$. If $\mu_1\mu_2\neq\mu_3\mu_4$, there exists no central configuration. If $\mu_1\mu_2=\mu_3\mu_4$, there is exactly one central configuration $(\theta_0, 5\pi/3-\theta_0, \theta_0, \pi/3-\theta_0)$ in the $1+4$ body problem, where $0<\theta_0\approx 0.6281<\pi/3$. Moreover,
$\frac{\mu_1}{\mu_3}=\frac{\mu_4}{\mu_2}=\frac{f(5\pi/3-\theta_0)}{f(\theta_0)}$.

When $f(\theta_1)=0$, we get the following Theorem:

\noindent{\bf Theorem 3.5.} The symmetric coorbital central configuration does not exist for system (2.4) with $f(\theta_1)=0$  and  $\theta_1=\theta_3$.

When $f(\theta_2)=0$, we get the following Theorem:

\noindent{\bf Theorem 3.6.}  Let $(\theta_1,\theta_2,\theta_3,\theta_4)$ be a coorbital central configuration solution of system (2.4).
Suppose that $\theta_1=\theta_3$, $f(\theta_2)=0$. If $\mu_1 \neq \mu_4$, there exists no central configuration. If $\mu_1=\mu_4$, there is exactly one central configuration $(\theta_0, \pi/3, \theta_0, 5\pi/3-2\theta_0)$ in the $1+4$ body problem, where $\pi/3<\theta_0\approx 1.4127<2\pi/3$. Moreover,
$(\mu_2+\mu_3)f(\theta_0)=\mu_1f(5\pi/3-2\theta_0)$.

When $f(\theta_4)=0$, we get the following Theorem:

\noindent{\bf Theorem 3.7.}  Let $(\theta_1,\theta_2,\theta_3,\theta_4)$ be a coorbital central configuration solution of system (2.4).
Suppose that $\theta_1=\theta_3$, $f(\theta_4)=0$. If $\mu_2 \neq \mu_3$, there exists no central configuration. If $\mu_2=\mu_3$, there is exactly one central configuration $(\theta_0, 5\pi/3-2\theta_0,\theta_0, \pi/3) $ in the $1+4$ body problem, where $\pi/3<\theta_0\approx 1.4127<2\pi/3$.

When $f(\theta_i)\neq 0, i=1,2,4$ and $f(\theta_1+\theta_2)\neq 0$, we get the following two Theorems:

\noindent{\bf Theorem 3.8.}  Let $(\theta_1,\theta_2,\theta_3,\theta_4)$ be a coorbital central configuration solution of system (2.4).
Under the above assumptions and also $\mu_1=\mu_4$, $\mu_2=\mu_3$, for each point in $F^{-1}(0)\cap
 (D_1\cup D_2\cup D_3)$ (see Figure 11), that is, the curve segments AC, DE, AB, and GH without the end-points, the $1+4$ bodies form a central configuration, where $F, D_1, D_2$ and $D_3$ are defined in (3.23).

\noindent{\bf Theorem 3.9.}  Let $(\theta_1,\theta_2,\theta_3,\theta_4)$ be a coorbital central configuration solution of system (2.4).
Under the above assumptions and $\mu_1 \neq \mu_4$, $\mu_2 \neq \mu_3$, also for each point in $F^{-1}(0)\cap
 (D_1\cup D_2\cup D_3)$ (see Figure 11), the $1+4$ bodies form a central configuration for suitable masses $\mu_1$, $\mu_2$, $\mu_3$ and $\mu_4$.

\section*{4. The proof of Theorem 3.2-3.7}
\setcounter{section}{4} \setcounter{equation}{0}

\subsection*{4.1. The proof of Theorem 3.2, 3.3 and 3.4.}
When $f(\theta_1+\theta_2)=0$, $\theta_1=\theta_3$, and $f(\theta_i)\neq 0, i=1,2,4$, system (3.1) reduces to
 \begin{equation}
\begin{aligned}
 \mu_2&f(\theta_1)=\mu_4f(\theta_4),\\
\mu_3&f(\theta_2)=\mu_1f(\theta_1),\\
\mu_4&f(\theta_1)=\mu_2f(\theta_2),\\
\mu_1&f(\theta_4)=\mu_3f(\theta_1),\\
\theta_1&+\theta_2+\theta_3+\theta_4=2\pi.\\
 \end{aligned}
 \end{equation}
By the first and the fourth equation of (4.1) we have $\mu_1\mu_2=\mu_3\mu_4$,
and by the second and the third equation of (4.1) we also get the same conclusion. This means $\mu_1\mu_2=\mu_3\mu_4$ is the necessary condition for the existence of the co-orbital central configuration under these assumptions.
Thus (4.1) is equivalent to
 \begin{equation}
\begin{aligned}
 \mu_2&f(\theta_1)=\mu_4f(\theta_4),\\
\mu_3&f(\theta_2)=\mu_1f(\theta_1),\\
\mu_1&\mu_2=\mu_3\mu_4,\\
\theta_1&+\theta_2+\theta_3+\theta_4=2\pi.\\
 \end{aligned}
 \end{equation}
 The system (4.2) above gives us
   \begin{equation}
   f^2(\theta_1)=f(\theta_2)f(\theta_4),
  \end{equation}
and the sign of $f(\theta_i)$ must be the same for all $i = 1,2,4$.

 By  $f(\theta_1+\theta_2)=0$, we have $\theta_1+\theta_2=\pi/3,\pi$ or $5\pi/3$. In the following we consider these three cases respectively.

 \begin{figure}[htb]
\begin{center}
\includegraphics[width=11cm,height=8cm]{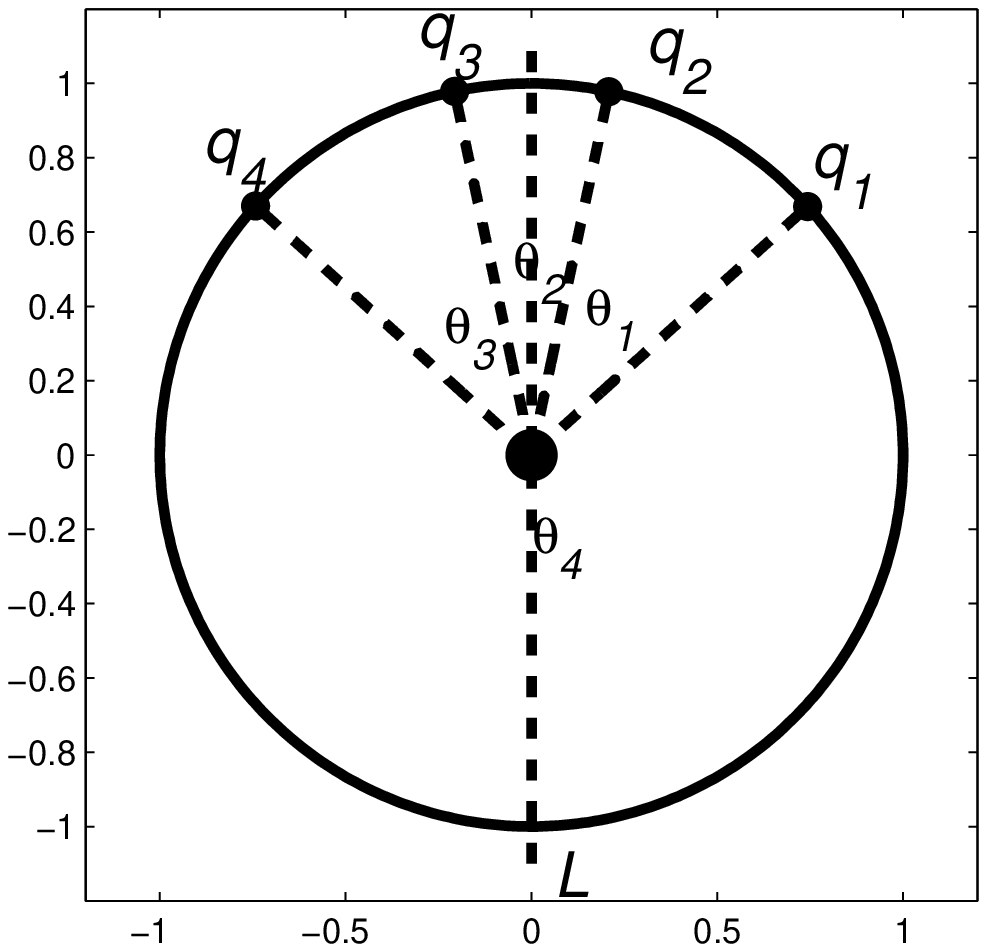}
\vspace{0cm}\caption{\label{TheLabel}\footnotesize The only coorbital central configuration according to Theorem 3.2}
\end{center}
\end{figure}

When $\theta_1+\theta_2=\pi/3$, let $\theta_1=\theta_3=\theta$, then $\theta_2=\pi/3-\theta$, $\theta_4=5\pi/3-\theta$, where $0<\theta<\pi/3$. Substituting these assumptions into (4.3) we have
 \begin{equation}
F_1(\theta)=f^2(\theta)-f(\pi/3-\theta)f(5\pi/3-\theta)=0.
\end{equation}

For $0<\theta<\pi/3$, then $ f'(\theta)>0$, $f(\theta)<0$. So
 \begin{equation}
(f^2(\theta))'=2f(\theta)f'(\theta)<0.
\end{equation}

We are going to prove that $-f(\pi/3-\theta)f(5\pi/3-\theta)$ monotonically decreases in $0<\theta<\pi/3$. With simple computation
$$(-f(\pi/3-\theta)f(5\pi/3-\theta))'=f'(\pi/3-\theta)f(5\pi/3-\theta)+f(\pi/3-\theta)f'(5\pi/3-\theta).$$
We divide the proof into two cases.

Case 1: $0<\theta\leq 5\pi/3-\theta_l=\theta_c-\pi/3$.

When $0<\theta\leq 5\pi/3-\theta_l=\theta_c-\pi/3<\pi/3$, then $f(\pi/3-\theta)<0$, $f(5\pi/3-\theta)<0$, $f'(\pi/3-\theta)>0$, $f'(5\pi/3-\theta)\geq 0$. So
 \begin{equation}
f'(\pi/3-\theta)f(5\pi/3-\theta)+f(\pi/3-\theta)f'(5\pi/3-\theta)<0.
\end{equation}

Case 2: $5\pi/3-\theta_l=\theta_c-\pi/3<\theta<\pi/3$.

When $0<5\pi/3-\theta_l=\theta_c-\pi/3<\theta<\pi/3$, then
 $$3+cos(\pi/3-\theta)-2sin(\pi/3-\theta)>0, sin(\pi/3-\theta)>0, cos(\pi/3-\theta)>0.$$ So
$$\begin{aligned}
&f(\pi/3-\theta)+f'(\pi/3-\theta)\\
&=sin(\pi/3-\theta)+cos(\pi/3-\theta)+\frac{3+cos(\pi/3-\theta)-2sin(\pi/3-\theta)}{16sin^3((\pi/3-\theta)/2)}\\
&>0,
\end{aligned}$$
that is,
\begin{equation}
f'(\pi/3-\theta)>-f(\pi/3-\theta)>0.
\end{equation}
Consider the function $G_1(\theta)=f(5\pi/3-\theta)-f'(5\pi/3-\theta)$. It is easily computed that
$$G_1'(\theta)=f''(5\pi/3-\theta)-f'(5\pi/3-\theta)>0$$
for $f''(5\pi/3-\theta)>0$, $-f'(5\pi/3-\theta)>0$.
Then  $G_1(\theta)<G_1(\pi/3)=-\frac{23}{36}\sqrt{3}+\frac{2}{3}<0$. So
\begin{equation}
-f(5\pi/3-\theta)>-f'(5\pi/3-\theta)>0
\end{equation}
Hence, by (4.7) and (4.8) we again get
$$f'(\pi/3-\theta)f(5\pi/3-\theta)+f(\pi/3-\theta)f'(5\pi/3-\theta)<0.$$

Case 1, case 2 and (4.5) mean that $F_1(\theta)=f^2(\theta)-f(\pi/3-\theta)f(5\pi/3-\theta)$ monotonically decreases in $0<\theta<\pi/3$.
It is easy to see that $F_1(\theta)\rightarrow +\infty$ as $\theta\rightarrow 0^{+}$ and $F_1(\theta)\rightarrow -\infty$ as $\theta\rightarrow {\pi/3}^{-}$.
So there is exactly one solution $\theta_0$ of $F_1
(\theta)=f^2(\theta)-f(\pi/3-\theta)f(5\pi/3-\theta)=0$ in $(0,\pi/3)$. Then by (4.1), $\frac{\mu_1}{\mu_3}=\frac{\mu_4}{\mu_2}=\frac{f(\pi/3-\theta_0)}{f(\theta_0)}>0$. Theorem 3.2 is proved.

When $\theta_1+\theta_2=\pi$, let $\theta_1=\theta_3=\theta$, then $\theta_2=\theta_4=\pi-\theta$, where $0<\theta<\pi$. Substituting these assumptions into the equation (4.3) we have
$$F_2(\theta)=f^2(\theta)-f^2(\pi-\theta)=0,$$
i.e.
$$(sin(\theta)(1-\frac{1}{8sin^3(\theta/2)}))^2-(sin(\theta)(1-\frac{1}{8cos^3(\theta/2)}))^2=0.$$

 \begin{figure}[htb]
\begin{center}
\includegraphics[width=11cm,height=8cm]{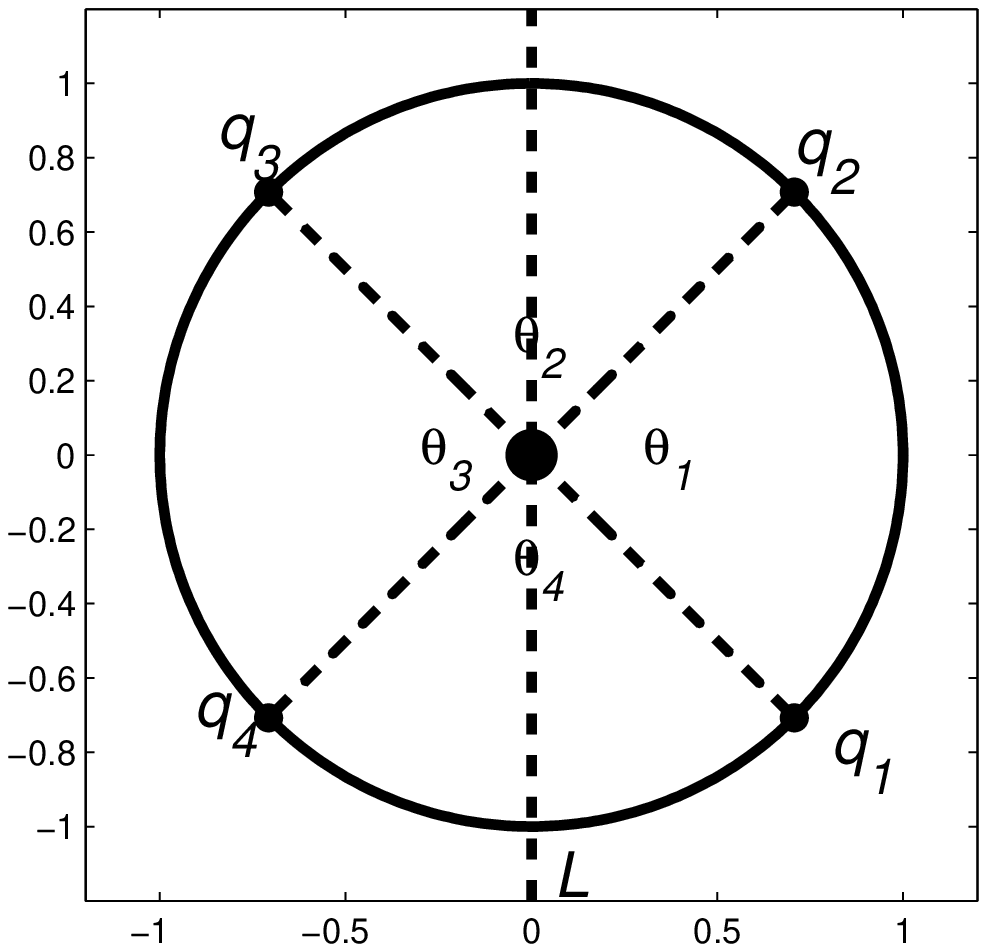}
\vspace{0cm}\caption{\label{TheLabel}\footnotesize The only coorbital central configuration according to Theorem 3.3}
\end{center}
\end{figure}

Since $sin(\theta) \neq 0$, the above equation is equivalent to
 \begin{equation}
|(1-\frac{1}{8sin^3(\theta/2)})|=|(1-\frac{1}{8cos^3(\theta/2)})|.
\end{equation}

If $(1-\frac{1}{8sin^3(\theta/2)})=(1-\frac{1}{8cos^3(\theta/2)})$, then $\theta=\pi/2$, and by (4.2) we have
 \begin{equation}
 \mu_1=\mu_3, \mu_4=\mu_2.
  \end{equation}

If
 $$(1-\frac{1}{8sin^3(\theta/2)})=-(1-\frac{1}{8cos^3(\theta/2)}),$$
 then
   \begin{equation}
   sin^{-3}(\theta/2)+ cos^{-3}(\theta/2)=16.
  \end{equation}
  Let $G_2(\theta)=sin^{-3}(\theta/2)+ cos^{-3}(\theta/2)-16$, then
  $$G_2'(\theta)=\frac{3}{2}sin^{-4}(\theta/2)cos^{-4}(\theta/2)(sin^3(\theta/2)- cos^3(\theta/2)).$$

  So if $\theta\in[0,\pi/2]$ then $G_2'(\theta)\leq 0$ and if $\theta\in[\pi/2,\pi]$ then $G_2'(\theta)\geq 0$.
  It is easy to see that $G_2(\theta)\rightarrow +\infty$ as $\theta\rightarrow 0^{+}$ and $\theta\rightarrow \pi^{-}$. By $G_2(\pi/2)=4\sqrt{2}-16<0$,
 there are exactly two solutions  of $F_2(\theta)=f^2(\theta)-f^2(\pi-\theta)=0$ in $(0,\pi)$.
  For
   $$G_2(\pi/3)=G_2(2\pi/3)=8(\frac{\sqrt{3}}{9}-1)<0,$$
  then one solution is in $(0,\pi/3)$ and the other solution is in $(2\pi/3,\pi)$.
 If the solution $\theta$ is in $(0,\pi/3)$, then $f(\theta_1)=f(\theta)<0$ and $f(\theta_4)=f(\pi-\theta)>0$.
If the solution $\theta$ is in $(2\pi/3,\pi)$, then $f(\theta_1)=f(\theta)>0$ and $f(\theta_4)=f(\pi-\theta)<0$. From the above two cases we have
$$\frac{\mu_4}{\mu_2}=\frac{f(\theta_1)}{f(\theta_4)}=\frac{f(\theta)}{f(\pi-\theta)}<0,$$
contradicting the fact that $\mu _i\in \mathbb{R}^+$. The Theorem 3.3 is proved.

\begin{figure}[htb]
\begin{center}
\includegraphics[width=11cm,height=8cm]{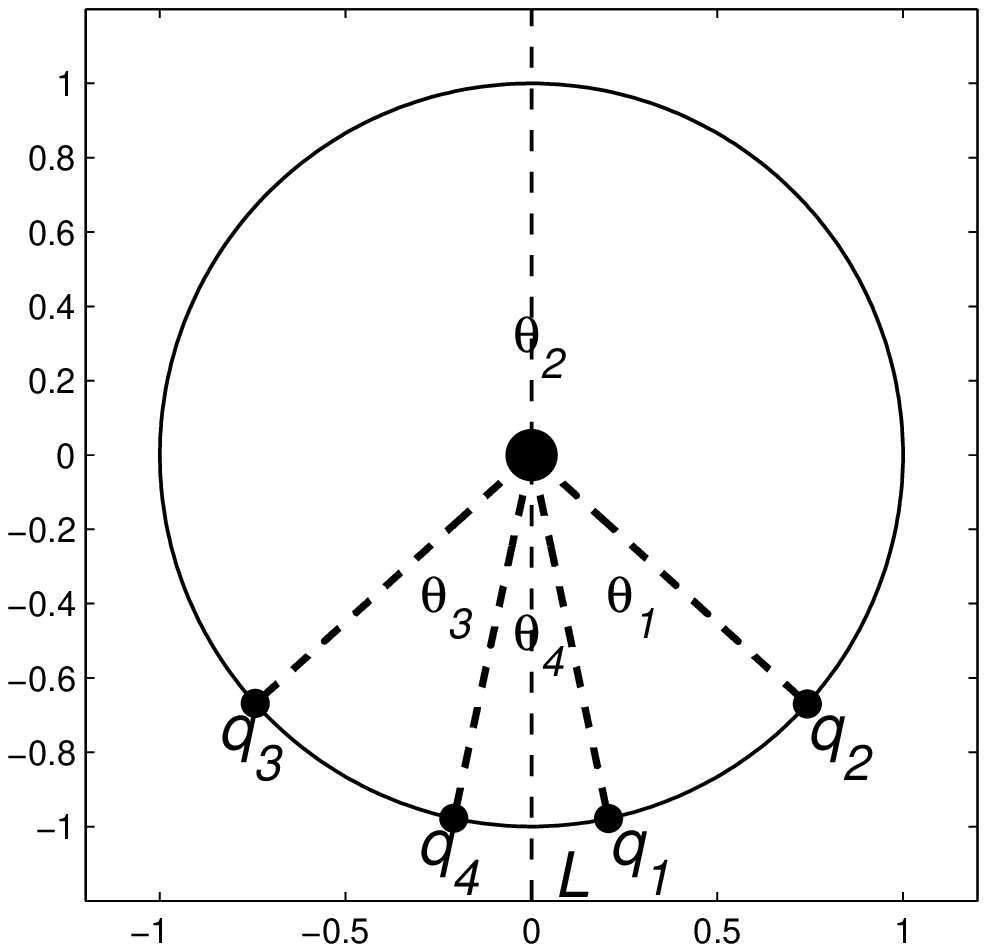}
\vspace{0cm}\caption{\label{TheLabel}\footnotesize The only coorbital central configuration according to Theorem 3.4}
\end{center}
\end{figure}

When $\theta_1+\theta_2=5\pi/3$, let $\theta_1=\theta_3=\theta$, then $\theta_2=5\pi/3-\theta$, $\theta_4=\pi/3-\theta$, where $0<\theta<\pi/3$. Substituting these assumptions into (4.3) we have
$$F_3(\theta)=f^2(\theta)-f(\pi/3-\theta)f(5\pi/3-\theta)=0,$$
which is the same with (4.4).  From the proof of Theorem 3.2, we find that there is exactly one solution $\theta_0$ for $F_3(\theta)=0$ in $(0,\pi/3)$. Then by (4.2), $\frac{\mu_1}{\mu_3}=\frac{\mu_4}{\mu_2}=\frac{f(5\pi/3-\theta_0)}{f(\theta_0)}>0$. The Theorem 3.4 is proved.

\subsection*{4.2. The proof of Theorem 3.5.}

When $f(\theta_1)=0$, $\theta_1=\theta_3$, system (3.1) becomes
 \begin{equation}
\begin{aligned}
 \mu_3&f(\theta_1+\theta_2)-\mu_4f(\theta_4)=0,\\
\mu_3&f(\theta_2)+\mu_4f(\theta_1+\theta_2)=0,\\
\mu_1&f(\theta_1+\theta_2)+\mu_2f(\theta_2)=0,\\
\mu_1&f(\theta_4)-\mu_2f(\theta_1+\theta_2)=0,\\
\theta_1&+\theta_2+\theta_3+\theta_4=2\pi.\\
 \end{aligned}
 \end{equation}
With simple calculation, system (4.12) is equivalent to

 \begin{equation}
\begin{aligned}
\mu_1&\mu_3=\mu_2\mu_4,\\
f(&\theta_2)f(\theta_4)+f^2(\theta_1+\theta_2)=0,\\
\mu_1&f(\theta_1+\theta_2)+\mu_2 f(\theta_2)=0,\\
\theta_1&+\theta_2+\theta_3+\theta_4=2\pi.\\
 \end{aligned}
 \end{equation}
For $f(\theta_1)=0$  and  $\theta_1=\theta_3$, we get $\theta_1=\theta_3=\pi/3$ and  $\theta_4=4\pi/3-\theta_2$ where $0<\theta_2<4\pi/3$.
The second equation of (4.13) becomes

\begin{equation}
f(\theta_2)f(4\pi/3-\theta_2)+f^2(\pi/3+\theta_2)=0.
 \end{equation}

We consider four subcases according to $\theta_2$ in $(0,\pi/3)$,$(\pi/3,\pi)$, $(\pi,4\pi/3)$ respectively or equal to $\pi/3$ or $\pi$.
Assume that $\theta_2\in (0,\pi/3)$, we have $4\pi/3-\theta_2 \in (\pi,4\pi/3)$, and then from the plot of $f$ we have that $f(\theta_2)<0$ and $f(4\pi/3-\theta_2)<0$. Now suppose that $\theta_2\in (\pi/3,\pi)$, then $4\pi/3-\theta_2 \in (\pi/3,\pi)$, similarly we have  $f(\theta_2)>0$ and $f(4\pi/3-\theta_2)>0$. When $\theta_2\in (\pi,4\pi/3)$, we have $4\pi/3-\theta_2 \in (0,\pi/3)$, this implies $f(\theta_2)<0$ and $f(4\pi/3-\theta_2)<0$. Finally when  $\theta_2=\pi/3$ or $\pi$, $f(\theta_2)f(4\pi/3-\theta_2)+f^2(\pi/3+\theta_2)=f^2(\pi/3+\theta_2)>0$.
The above analysis shows that (4.14) does not hold. The Theorem 3.5 is proved.

\subsection*{4.3. The proof Theorem 3.6 and 3.7.}

\begin{figure}[htb]
\begin{center}
\includegraphics[width=11cm,height=8cm]{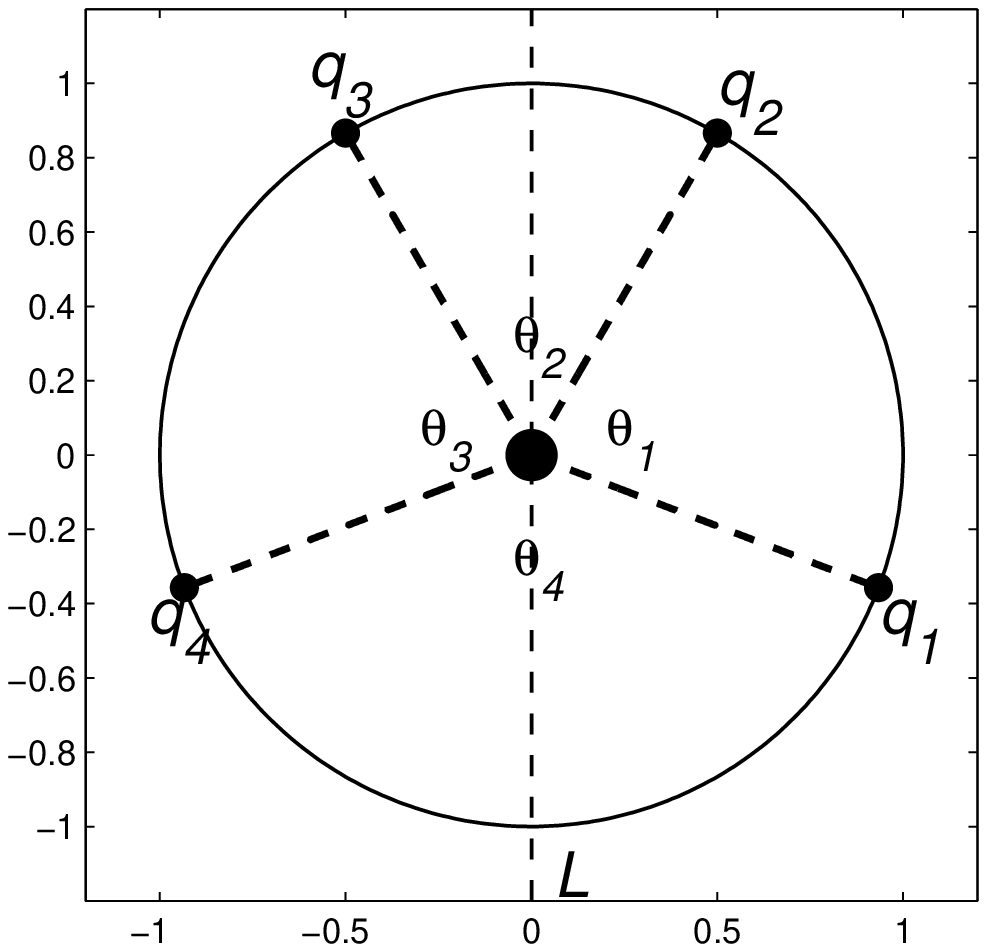}
\vspace{0cm}\caption{\label{TheLabel}\footnotesize The only coorbital central configuration according to Theorem 3.6}
\end{center}
\end{figure}

When $f(\theta_2)=0$, $\theta_1=\theta_3$, system (3.1) becomes
  \begin{equation}
\begin{aligned}
 \mu_2&f(\theta_1)+\mu_3f(\theta_1+\theta_2)-\mu_4f(\theta_4)=0,\\
\mu_4&f(\theta_1+\theta_2)-\mu_1f(\theta_1)=0,\\
\mu_4&f(\theta_1)-\mu_1f(\theta_1+\theta_2)=0,\\
\mu_1&f(\theta_4)-\mu_2f(\theta_1+\theta_2)-\mu_3f(\theta_1)=0,\\
\theta_1&+\theta_2+\theta_3+\theta_4=2\pi,\\
 \end{aligned}
 \end{equation}

From the second and the third equation we obtain that $\mu_1=\mu_4$ and $f(\theta_1+\theta_2)=f(\theta_1)$.
This means $\mu_1=\mu_4$ is the necessary condition for the existence of the co-orbital central configuration under these assumptions.
Then system (4.15) is equivalent to
  \begin{equation}
\begin{aligned}
 \mu&_1=\mu_4,\\
f(&\theta_1+\theta_2)=f(\theta_1),\\
(\mu&_2+\mu_3)f(\theta_1)-\mu_1f(\theta_4)=0,\\
\theta_1&+\theta_2+\theta_3+\theta_4=2\pi,\\
 \end{aligned}
 \end{equation}

 $f(\theta_2)=0$ implies that $\theta_2=\pi/3, \pi$ or $5\pi/3$.

 Assume that $\theta_2=\pi/3$, then $\theta_4=5\pi/3-2\theta_1$ where $0<\theta_1<5\pi/6$ and $\theta_1\neq \pi/3$ for $f(\theta_1)\neq 0$.
The third equation of (4.16) means that $f(\theta_1)$ and $f(\theta_4)$ have the same sign. When $f(\theta_1)>0$ and $f(\theta_4)>0$ we have
$\pi/3<\theta_1<2\pi/3$. When $f(\theta_1)<0$ and $f(\theta_4)<0$ we have $0<\theta_1<\pi/3$. Then we consider two subcases according to $\theta_1$ in $(0,\pi/3)$ or $(\pi/3,2\pi/3)$. Hence we must solve
$$f(\pi/3+\theta_1)-f(\theta_1)=0 \text{ for }\theta\in(0,\pi/3)\cup(\pi/3,2\pi/3).$$
Clearly $f'(\pi/3+\theta_1)-f'(\theta_1)<0$ in $(0,2\pi/3)$ due to $f''(x)<0$ in $(0,\pi)$.

It is easy to see that $f(\pi/3+\theta_1)-f(\theta_1)\rightarrow +\infty$ as $\theta_1\rightarrow 0^{+}$ , $f(\pi/3+\theta_1)-f(\theta_1)\rightarrow f(2\pi/3)>0$ as $\theta_1\rightarrow \pi/3$, and
$f(\pi/3+\theta_1)-f(\theta_1)\rightarrow -f(2\pi/3)<0$ as $\theta_1\rightarrow {2\pi/3}^{-}$.
So $f(\pi/3+\theta_1)-f(\theta_1)=0$ must have exactly one root in $(\pi/3,2\pi/3)$.

Suppose that $\theta_2=\pi$, then $\theta_4=\pi-2\theta_1$ where $0<\theta_1<\pi/2$. When $\theta_1\in (0,\pi/3)$, then $\theta_4=\pi-2\theta_1 \in (\pi/3,\pi)$, which is a contradiction with the fact that $f(\theta_1)$ and $f(\theta_4)$ must have the same sign.  When $\theta_1\in (\pi/3,\pi/2)$, then $\theta_4=\pi-2\theta_1 \in (0,\pi/3)$, again we have the same contradiction.

Now consider $\theta_2=5\pi/3$, then $\theta_4=\pi/3-2\theta_1$ where $0<\theta_1<\pi/6$. From the plot of $f$ we have $f(\theta_1)<0$ and $f(\theta_1+\theta_2)=f(\theta_1+5\pi/3)>0$. Then the second equation of (4.16) does not hold. We complete the proof of Theorem 3.6.

The proof of Theorem 3.7 is in a similar way of the above.

\begin{figure}[htb]
\begin{center}
\includegraphics[width=11cm,height=8cm]{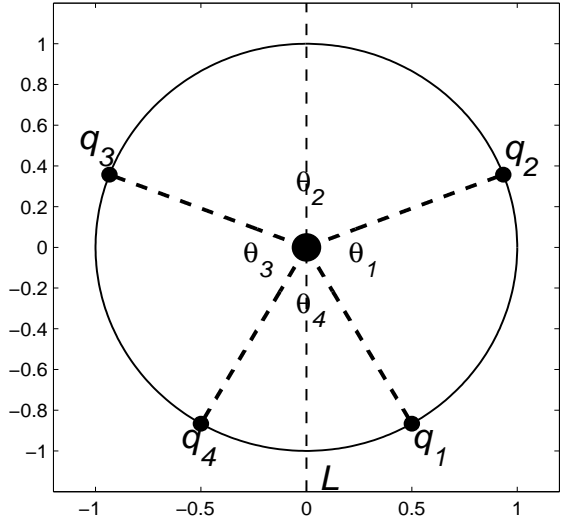}
\vspace{0cm}\caption{\label{TheLabel}\footnotesize The only coorbital central configuration according to Theorem 3.7}
\end{center}
\end{figure}

\subsection*{4.4. The numerical evidences of Theorem 3.8 and 3.9.}

Now we consider system (3.1) with $f(\theta_i)\neq 0, i=1,2,4$ and $f(\theta_1+\theta_2)\neq 0$.
With simple computation, by the first two equations and the second two equations of (3.1) respectively we have
  \begin{equation}
\begin{aligned}
f(\theta_1)&=\frac{{\mu}^2_4f(\theta_4)+{\mu}^2_3f(\theta_2)}{\mu_1\mu_3+\mu_2\mu_4},\\
f(\theta_1+\theta_2)&=\frac{\mu_1\mu_4f(\theta_4)-\mu_2\mu_3f(\theta_2)}{\mu_1\mu_3+\mu_2\mu_4},\\
 \end{aligned}
 \end{equation}
and
  \begin{equation}
\begin{aligned}
f(\theta_1)&=\frac{{\mu}^2_2f(\theta_2)+{\mu}^2_1f(\theta_4)}{\mu_1\mu_3+\mu_2\mu_4},\\
f(\theta_1+\theta_2)&=\frac{\mu_1\mu_4f(\theta_4)-\mu_2\mu_3f(\theta_2)}{\mu_1\mu_3+\mu_2\mu_4}.\\
 \end{aligned}
 \end{equation}
The second equation in (4.17) and (4.18) are the same, and the first equation in (4.17) and (4.18) give us
$$({\mu}^2_3-{\mu}^2_2)f(\theta_2)=({\mu}^2_1-{\mu}^2_4)f(\theta_4).$$
Then the system (3.1) reduces to
  \begin{equation}
\begin{aligned}
f&(\theta_1)=\frac{{\mu}^2_4f(\theta_4)+{\mu}^2_3f(\theta_2)}{\mu_1\mu_3+\mu_2\mu_4},\\
f&(\theta_1+\theta_2)=\frac{\mu_1\mu_4f(\theta_4)-\mu_2\mu_3f(\theta_2)}{\mu_1\mu_3+\mu_2\mu_4},\\
(&{\mu}^2_3-{\mu}^2_2)f(\theta_2)=({\mu}^2_1-{\mu}^2_4)f(\theta_4),\\
\theta&_1+\theta_2+\theta_3+\theta_4=2\pi.
 \end{aligned}
 \end{equation}

In the following we consider two cases $({\mu}^2_3-{\mu}^2_2)f(\theta_2)=({\mu}^2_1-{\mu}^2_4)f(\theta_4)=0$ and $({\mu}^2_3-{\mu}^2_2)f(\theta_2)=({\mu}^2_1-{\mu}^2_4)f(\theta_4)\neq 0$ respectively.

\begin{figure}[htb]
\begin{center}
\includegraphics[width=10cm,height=7cm]{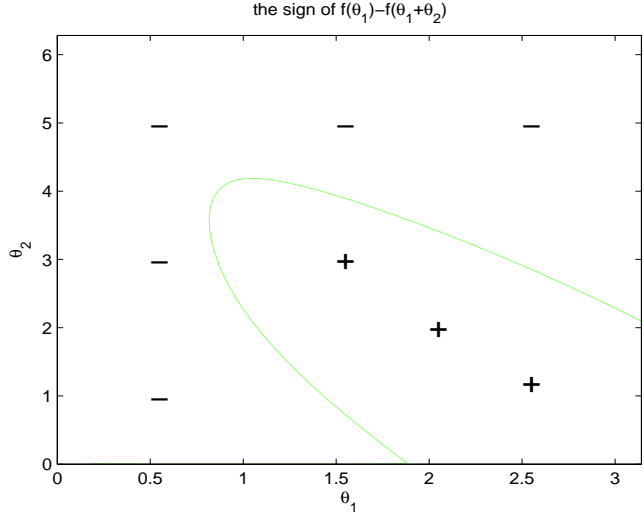}
\vspace{0cm}\caption{\label{TheLabel}\footnotesize The sign of $f(\theta_1)-f(\theta_1+\theta_2)$ and the curve $f(\theta_1)-f(\theta_1+\theta_2)=0$}
\end{center}
\end{figure}

When $({\mu}^2_3-{\mu}^2_2)f(\theta_2)=({\mu}^2_1-{\mu}^2_4)f(\theta_4)=0$, with the assumptions that $f(\theta_i)\neq 0, i=1,2,4$ and $f(\theta_1+\theta_2)\neq 0$, we have $ \mu_1=\mu_4,\mu_2=\mu_3$, and then the system (4.19) reduces to
  \begin{equation}
\begin{aligned}
f&(\theta_1)=\frac{{\mu}^2_1f(\theta_4)+{\mu}^2_2f(\theta_2)}{2\mu_1\mu_2},\\
f&(\theta_1+\theta_2)=\frac{{\mu}^2_1f(\theta_4)-{\mu}^2_2f(\theta_2)}{2\mu_1\mu_2},\\
\theta&_1+\theta_2+\theta_3+\theta_4=2\pi.
 \end{aligned}
 \end{equation}
Let $\lambda=\frac{\mu_1}{\mu_2}$, the first two equations of (4.20) give us
$$f(\theta_1)+f(\theta_1+\theta_2)=\lambda f(\theta_4),$$
$$f(\theta_1)-f(\theta_1+\theta_2)=\frac{1}{\lambda }f(\theta_2).$$
It follows that
  \begin{equation}
\begin{aligned}
f^2(\theta_1)-f^2(\theta_1+\theta_2)&=f(\theta_2)f(\theta_4),\\
\frac{f(\theta_2)}{f(\theta_1)-f(\theta_1+\theta_2)}&=\lambda>0,\\
 \end{aligned}
 \end{equation}
where $\theta_4=2\pi-2\theta_1-\theta_2$, and $0<\theta_1<\pi, 0<2\theta_1+\theta_2<2\pi$.

\begin{figure}[htb]
\begin{center}
\includegraphics[width=10cm,height=7cm]{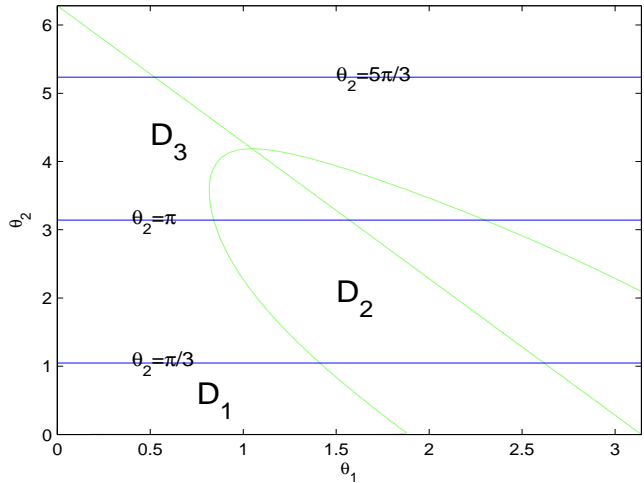}
\vspace{0cm}\caption{\label{TheLabel}\footnotesize The regions $D_1$,$D_2$ and $D_3$}
\end{center}
\end{figure}
Let
  \begin{equation}
\begin{aligned}
F&(\theta_1,\theta_2)=f^2(\theta_1)-f^2(\theta_1+\theta_2)-f(\theta_2)f(2\pi-2\theta_1-\theta_2),\\
D&_1=\{(\theta_1,\theta_2):0<\theta_2<\frac{\pi}{3}, f(\theta_1)-f(\theta_1+\theta_2)<0, 0<\theta_1<\pi, 0<2\theta_1+\theta_2<2\pi\},\\
D&_2=\{(\theta_1,\theta_2):\frac{\pi}{3}<\theta_2<\pi, f(\theta_1)-f(\theta_1+\theta_2)>0, 0<\theta_1<\pi, 0<2\theta_1+\theta_2<2\pi\},\\
D&_3=\{(\theta_1,\theta_2):\pi<\theta_2<\frac{5\pi}{3}, f(\theta_1)-f(\theta_1+\theta_2)<0, 0<\theta_1<\pi, 0<2\theta_1+\theta_2<2\pi\},\\
D&_4=\{(\theta_1,\theta_2):\frac{5\pi}{3}<\theta_2<2\pi, f(\theta_1)-f(\theta_1+\theta_2)>0, 0<\theta_1<\pi, 0<2\theta_1+\theta_2<2\pi\}.\\
 \end{aligned}
 \end{equation}
When $\frac{5\pi}{3}<\theta_2<2\pi$, we have $\frac{5\pi}{3}<\theta_1+\theta_2<2\pi$ and $\theta_1<\pi/3$, thus $f(\theta_1)-f(\theta_1+\theta_2)<0$. This means $D_4$ is an empty set. Obviously the signs of $f(\theta_2)$ and $f(\theta_1)-f(\theta_1+\theta_2)$ in $D_1$, $D_2$ and $D_3$ are the same, thus the mass ratio $\lambda=\frac{\mu_1}{\mu_2}$ is guaranteed to be positive.

\begin{figure}[htb]
\begin{center}
\includegraphics[width=17cm,height=10cm]{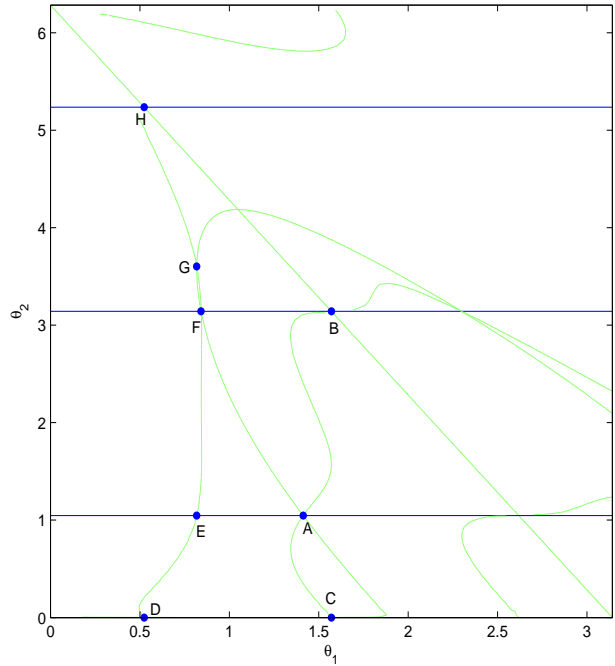}
\vspace{0cm}\caption{\label{TheLabel}\footnotesize The curves $F(\theta_1,\theta_2)=0$ in $D_1$,$D_2$ and $D_3$: the curve segments AC, DE, AB, and GH without the end-points}
\end{center}
\end{figure}

\begin{figure}[htb]
\begin{center}
\includegraphics[width=17cm,height=8cm]{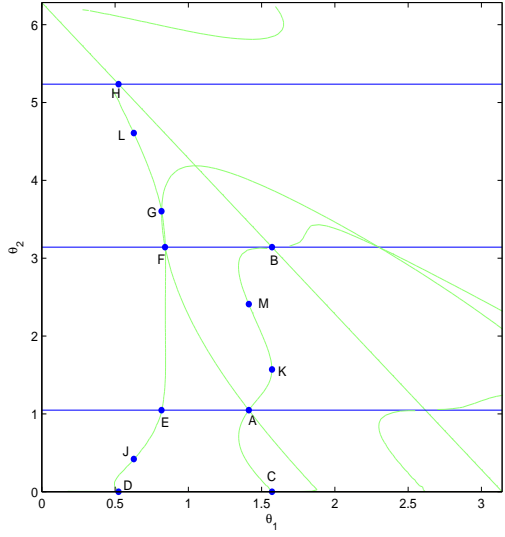}
\vspace{0cm}\caption{\label{TheLabel}\footnotesize Special points $J$, $K$,$L$ and $M$, which correspond to the coorbital central configurations of Theorem $3.2$, $3.3$, $3.4$ and $3.7$}
\end{center}
\end{figure}

\begin{table}
\caption{Some numerical results $\theta_1$, $\theta_2$ along $F^{-1}(0)\cap D_1$  and related mass ratios $\lambda$}
\begin{center}
\begin{tabu} to 0.8\textwidth{X[c]|X[c]|X[c]|X[c]|X[c]|X[c]}
\hline
$\theta_1$  &$\theta_2$              &$\lambda$      &$\theta_1$  &$\theta_2$              &$\lambda$ \\
\hline
0.4922 & 0.1 & 73.2972&1.5212 &0.1& 1.8820e+03\\
0.5199 &0.2&12.8052&1.4739&0.2&234.2838\\
0.5692&0.3&5.4154&1.4312&0.3&69.1611\\
0.6190&0.4&3.0292&1.3951&0.4&29.1372\\
0.6647&0.5&1.8989&1.3675&0.5&14.9637\\
0.7053&0.6&1.2437&1.3498&0.6&8.7330\\
0.7404&0.7&0.8125&1.3432&0.7&5.5832\\
0.7695&0.8&0.5023&1.3486&0.8&3.8296\\
0.7927&0.9&0.2653&1.3662&0.9&2.7810\\
0.8102&1.0&0.0766&1.3954&1.0&2.1227\\
\hline
\end{tabu}
\end{center}
\end{table}

\begin{table}
\caption{Some numerical results $\theta_1$, $\theta_2$ along $F^{-1}(0)\cap D_2$  and related mass ratio $\lambda$}
\begin{center}
\begin{tabu} to 0.8\textwidth{X[c]|X[c]|X[c]|X[c]|X[c]|X[c]}
\hline
$\theta_1$  &$\theta_2$              &$\lambda$      &$\theta_1$  &$\theta_2$              &$\lambda$ \\
\hline
1.4341&1.1&1.6921&1.4608&2.2&0.6637\\
1.4779&1.2&1.4123&1.4375&2.3&0.6003\\
1.5196&1.3&1.2310&1.4150&2.4&0.5343\\
1.5511&1.4&1.1158&1.3938&2.5&0.4661\\
1.5677&1.5&1.0406&1.3748&2.6&0.3959\\
1.5703&1.6&0.9848&1.3588&2.7&0.3241\\
1.5626&1.7&0.9355&1.3473&2.8&0.2509\\
1.5479&1.8&0.8868&1.3432&2.9&0.1768\\
1.5289&1.9&0.8357&1.3528&3.0&0.1024\\
1.5073&2.0&0.7816&1.4020&3.1&0.0292\\
1.4843&2.1&0.7242&  &  &   \\
\hline
\end{tabu}
\end{center}
\end{table}

\begin{table}[!htbp]
\caption{Some numerical results $\theta_1$, $\theta_2$ along $F^{-1}(0)\cap D_3$  and related mass ratio $\lambda$}
\begin{center}
\begin{tabu} to 0.8\textwidth{X[c]|X[c]|X[c]|X[c]|X[c]|X[c]}
\hline
$\theta_1$  &$\theta_2$              &$\lambda$      &$\theta_1$  &$\theta_2$              &$\lambda$ \\
\hline
0.8058&3.7&7.9282&0.6539&4.5&0.4723\\
0.7924&3.8&3.7257&0.6300&4.6&0.3704\\
0.7769&3.9&2.3431&0.6055&4.7&0.2855\\
0.7597&4.0&1.6559&0.5805&4.8&0.2135\\
0.7408&4.1&1.2385&0.5553&4.9&0.1519\\
0.7207&4.2&0.9579&0.5303&5.0&0.0988\\
0.6994&4.3&0.7535&0.5071&5.1&0.0530\\
0.6771&4.4&0.5969&0.4922&5.2&0.0132\\
\hline
\end{tabu}
\end{center}
\end{table}

The sign of $f(\theta_1)-f(\theta_1+\theta_2)$ is determined in Figure 9. The regions $D_1$, $D_2$ and $D_3$ can be seen in Figure 10, and the curve of $F(\theta_1,\theta_2)=0$ is plotted in Figure 11. With simple computation we have the coordinates of these intersection points: $A(1.4127,\frac{\pi}{3})$, $B(\frac{\pi}{2},\pi)$, $C(\frac{\pi}{2},0)$, $D(\frac{\pi}{6},0)$, $E(0.8167,\frac{\pi}{3})$, $F(0.8413,\pi)$, $G(0.8167,3.6026)$, $H(\frac{\pi}{6},\frac{5\pi}{3})$, where point $A$ corresponds to  the coorbital central configuration of Theorem $3.6$. Also, we get the other four points in $F^{-1}(0)\cap(D_1\cup D_2\cup D_3)$: $J(0.6281,0.4191)$, $K(\frac{\pi}{2},\frac{\pi}{2})$, $L(0.6281,4.6079)$, $M(1.4127,2.4106)$ (see Figure 12), which correspond to the coorbital central configurations of Theorem $3.2$, $3.3$, $3.4$ and $3.7$. In Table 1 we give some numerical results
along $F^{-1}(0)\cap D_1$  and find that there are two values on $\theta_1$ corresponding to any $\theta_2 \in (0,\frac{\pi}{3})$. Also we give some numerical results along $F^{-1}(0)\cap D_2$ and $F^{-1}(0)\cap D_3$ in Table 2 and Table 3 respectively.

When $({\mu}^2_3-{\mu}^2_2)f(\theta_2)=({\mu}^2_1-{\mu}^2_4)f(\theta_4) \neq 0$, with the assumptions that $f(\theta_i)\neq 0, i=1,2,4$ and $f(\theta_1+\theta_2)\neq 0$, we have $ \mu_1 \neq \mu_4,\mu_2 \neq \mu_3$, and
\begin{equation}
f(\theta_2)=\frac{{\mu}^2_1-{\mu}^2_4}{{\mu}^2_3-{\mu}^2_2}f(\theta_4).
\end{equation}
Substituting the above equation into the the first two equations of (4.19), we have
\begin{equation}
f(\theta_1)=\frac{\mu_1\mu_3-\mu_2\mu_4}{{\mu}^2_3-{\mu}^2_2}f(\theta_4),
\end{equation}
\begin{equation}
f(\theta_1+\theta_2)=\frac{\mu_3\mu_4-\mu_1\mu_2}{{\mu}^2_3-{\mu}^2_2}f(\theta_4).
\end{equation}
Then,
$$f^2(\theta_1)-f^2(\theta_1+\theta_2)=\frac{{\mu}^2_1-{\mu}^2_4}{{\mu}^2_3-{\mu}^2_2}f^2(\theta_4),$$

\begin{table}
\caption{Some numerical results $\theta_1$, $\theta_2$ along $F^{-1}(0)\cap D_1$  and related conditions of these masses}
\begin{center}
\begin{tabu} to 1.1\textwidth{X[c]|X[c]|X[c]|X[c]|X[c]|X[c]|X[c]|X[c]}
\hline
$\theta_1$  &$\theta_2$  &$\frac{\mu_1+\mu_4}{\mu_2+\mu_3}$   &$\frac{\mu_1-\mu_4}{\mu_3-\mu_2}$         &$\theta_1$  &$\theta_2$   &$\frac{\mu_1+\mu_4}{\mu_2+\mu_3}$   &$\frac{\mu_1-\mu_4}{\mu_3-\mu_2}$\\
\hline
0.4922 & 0.1 & 73.2972&17.0322& 1.5212 &0.1& 1.8820e+03&-77.5064\\
0.5199 &0.2&12.8052&5.6441 &1.4739&0.2&234.2838&-19.3970\\
0.5692&0.3&5.4154&3.5674& 1.4312&0.3&69.1611&-8.5529\\
0.6190&0.4&3.0292&2.8349&1.3951&0.4&29.1372&-4.6917\\
0.6647&0.5&1.8989&2.5298&1.3675&0.5&14.9637&-2.8510\\
0.7053&0.6&1.2437&2.4327&1.3498&0.6&8.7330&-1.8086\\
0.7404&0.7&0.8125&2.4765&1.3432&0.7&5.5832&-1.1469\\
0.7695&0.8&0.5023&2.6403&1.3486&0.8&3.8296&-0.6917\\
0.7927&0.9&0.2653&2.9459&1.3662&0.9&2.7810&-0.3594\\
0.8102&1.0&0.0766&3.4274&1.3954&1.0&2.1227&-0.1033\\

\hline
\end{tabu}
\end{center}
\end{table}

\begin{table}
\caption{Some numerical results $\theta_1$, $\theta_2$ along $F^{-1}(0)\cap D_2$  and related conditions of these masses}
\begin{center}
\begin{tabu} to 0.9\textwidth{X[c]|X[c]|X[c]|X[c]|X[c]|X[c]|X[c]|X[c]}
\hline
$\theta_1$  &$\theta_2$  &$\frac{\mu_1+\mu_4}{\mu_2+\mu_3}$   &$\frac{\mu_1-\mu_4}{\mu_3-\mu_2}$         &$\theta_1$  &$\theta_2$   &$\frac{\mu_1+\mu_4}{\mu_2+\mu_3}$   &$\frac{\mu_1-\mu_4}{\mu_3-\mu_2}$\\
\hline
1.4341&1.1&1.6921&0.1065&1.4608&2.2&0.6637&4.4956\\
1.4779&1.2&1.4123&0.2922&1.4375&2.3&0.6003&8.1788\\
1.5196&1.3&1.2310&0.4710&1.4150&2.4&0.5343&79.7431\\
1.5511&1.4&1.1158&0.6551&1.3938&2.5&0.4661&-8.7964\\
1.5677&1.5&1.0406&0.8512&1.3748&2.6&0.3959&-3.7527\\
1.5703&1.6&0.9848& 1.0645&1.3588&2.7&0.3241&-2.1595\\
1.5626&1.7&0.9355&1.3035&1.3473&2.8&0.2509&-1.3575\\
1.5479&1.8&0.8868&1.5839&1.3432&2.9&0.1768&-0.8588\\
1.5289&1.9&0.8357&1.9341&1.3528&3.0&0.1024&-0.5000\\
1.5073&2.0&0.7816&2.4105&1.4020&3.1&0.0292&-0.1902\\
1.4843&2.1&0.7242&3.1410&  &  &  & \\

\hline
\end{tabu}
\end{center}
\end{table}

\begin{table}[!htbp]
\caption{Some numerical results $\theta_1$, $\theta_2$ along $F^{-1}(0)\cap D_3$  and related conditions of these masses}
\begin{center}
\begin{tabu} to 0.9\textwidth{X[c]|X[c]|X[c]|X[c]|X[c]|X[c]|X[c]|X[c]}
\hline
$\theta_1$  &$\theta_2$  &$\frac{\mu_1+\mu_4}{\mu_2+\mu_3}$   &$\frac{\mu_1-\mu_4}{\mu_3-\mu_2}$         &$\theta_1$  &$\theta_2$   &$\frac{\mu_1+\mu_4}{\mu_2+\mu_3}$   &$\frac{\mu_1-\mu_4}{\mu_3-\mu_2}$\\
\hline
0.8058&3.7&7.9282&0.3052&0.6539&4.5&0.4723&0.3874\\
0.7924&3.8&3.7257&0.3402&0.6300&4.6&0.3704&0.3652\\
0.7769&3.9&2.3431&0.3684&0.6055&4.7&0.2855&0.3359\\
0.7597&4.0&1.6559&0.3897&0.5805&4.8&0.2135&0.2992\\
0.7408&4.1&1.2385&0.4036&0.5553&4.9&0.1519&0.2552\\
0.7207&4.2&0.9579&0.4105&0.5303&5.0&0.0988&0.2028\\
0.6994&4.3&0.7535&0.4101&0.5071&5.1&0.0530&0.1404\\
0.6771&4.4&0.5969&0.4024&0.4922&5.2&0.0132&0.0573\\

\hline
\end{tabu}
\end{center}
\end{table}

combining with (4.23), again we have
 \begin{equation}
f^2(\theta_1)-f^2(\theta_1+\theta_2)=f(\theta_2)f(\theta_4).
 \end{equation}
Also from (4.24) and (4.25), we have
$$f(\theta_1)+f(\theta_1+\theta_2)=\frac{\mu_1+\mu_4}{\mu_2+\mu_3}f(\theta_4),$$
$$f(\theta_1)-f(\theta_1+\theta_2)=\frac{\mu_1-\mu_4}{\mu_3-\mu_2}f(\theta_4).$$
Then system (4.19) reduces to
 \begin{equation}
\begin{aligned}
f(\theta_1)+f(\theta_1+\theta_2)=\frac{\mu_1+\mu_4}{\mu_2+\mu_3}f(\theta_4),\\
f(\theta_1)-f(\theta_1+\theta_2)=\frac{\mu_1-\mu_4}{\mu_3-\mu_2}f(\theta_4),\\
f^2(\theta_1)-f^2(\theta_1+\theta_2)=f(\theta_2)f(\theta_4),\\
 \end{aligned}
 \end{equation}
where $\theta_4=2\pi-2\theta_1-\theta_2$, and $0<\theta_1<\pi, 0<2\theta_1+\theta_2<2\pi$.

For $\frac{\mu_1+\mu_4}{\mu_2+\mu_3}>0$, $f(\theta_1)+f(\theta_1+\theta_2)$ and $f(\theta_4)$ should have the same sign, which is equivalent to the fact that
$f(\theta_1)-f(\theta_1+\theta_2)$ and $f(\theta_2)$ have the same sign for $f^2(\theta_1)-f^2(\theta_1+\theta_2)=f(\theta_2)f(\theta_4)$. Again we get the curves $F(\theta_1,\theta_2)=0$ in $D_1$,$D_2$ and $D_3$, that the curve segments AC, DE, AB, and GH without the end-points (see in Figure 11). In Table 4, 5 and 6 we give some numerical results along
$F^{-1}(0)\cap D_1$, $F^{-1}(0)\cap D_2$ and $F^{-1}(0)\cap D_3$ respectively, where $\mu_1 \neq \mu_4$, $\mu_2 \neq \mu_3$.
Thus we complete the proof of Theorem 3.8 and 3.9.
\section*{Acknowledgements}
The authors are Supported by Natural Science Foundation of China (NFSC11703006) and
the Scientific Research Foundation of Huaiyin Institute of
Technology.

\end{document}